\documentclass[prb,showpacs,twocolumn]{revtex4}%
\usepackage{amsfonts}
\usepackage{amsmath}
\usepackage{amssymb}
\usepackage{graphicx}%
\begin{document}
\title{Equation of state for shock-compressed porous molybdenum from first-principles
mean-field potential calculations}
\author{Qili Zhang, Ping Zhang, Gongmu Zhang, Haifeng Liu}
\affiliation{Institute of Applied Physics and Computational Mathematics, P.O. Box 8009,
Beijing 100088, People's Republic of China}

\begin{abstract}
The Hugoniot curves for shock-compressed molybdenum with initial
porosities of 1.0, 1.26, 1.83, and 2.31 are theoretically
investigated. The method of calculations combines the
first-principles treatment for zero- and finite-temperature
electronic contribution and the mean-field-potential approach for
the ion-thermal contribution to the total free energy. Our
calculated results reproduce the Hugoniot properties of porous
molybdenum quite well. At low porosity, in particular, the
calculations show a complete agreement with the experimental
measurements over the full range of data. For the two large porosity
values of 1.83 and 2.31, our results are well in accord with the
experimental data points up to the particle velocity of 3.5 km/s,
and tend to overestimate the shock-wave velocity and Hugoniot
pressure when further increasing the particle velocity. In addition,
the temperature along the principal Hugoniot is also extensively
investigated for porous molybdenum.

\end{abstract}
\maketitle

Molybdenum (Mo) is a high technology metal with wide engineering applications
for its thermal and mechanical strength, and also chemical resistances. It has
been used as flyer and/or back reflector for shock wave experiments to provide
accurate standard of high-pressure equation of state (EOS) \cite{Carter,Mao}.
From this view, the precise EOS of Mo is critical for theoretical and
practical implications. To obtain the EOS of Mo, the shock-compressed crystal
Mo has been investigated by using different experiment systems
\cite{Walsh,McQueen,Krupnikov,McQueen1,Ragan,Marsh,Al,
Ragan1,Ragan2,Mitchell,Hixson,Trunin2}. The EOS shock-wave data have now been
obtained at pressures ranging from 0.1 GPa up to a few TPa. Furthermore, to
extend the EOS to regions of higher internal energy and temperature, the
porous samples of Mo have also been experimentally investigated by Bakanova
\textit{et al}. \cite{Bakanova} and Trunin \textit{et al}. \cite{Trunin1}.
Compared to the extensive theoretical calculations and analysis of the
shock-compressed crystal Mo, the EOS shock-wave data of the \textit{porous} Mo
remain yet to be theoretically exploited and understood, which is a main
driving force for our present study.

In this paper we calculate the Hugoniots of porous Mo with experimentally
relevant \cite{Trunin1,Bakanova} porosities $m$=1.0, 1.26, 1.83, and 2.31
using the mean-field potential (MFP) approach facilitated with the
first-principles calculations. The MFP approach, which will be briefly
described below, was initiated by Wang and co-workers \cite{Wang}, and proved
to provide a numerically convenient way to take into account the ion-thermal
contribution in the total internal energy of the metal based on the
first-principles calculation of zero-temperature internal energy. A lot of
elemental metals have been successfully tested
\cite{Wang,Wang1,Wang,Li,Wang3,Wang4,Wang5}, and a general agreement between
the MFP calculations and the experimental Hugoniots has been achieved under
low shock temperatures (compared to the Fermi temperature $T_{F}$ of the
valence electrons) and low Hugoniot pressures (typically within 0.1 TPa).
Also, the shock-compressed porous carbon has been investigated using this
approach \cite{Wang7}. Our present results of the Hugoniots for the porous Mo
with different porosities show good agreement with the attainable experimental
data from several groups.

Now we start with a brief review of the MFP approach. For a system with a
given averaged atomic volume $V$ and temperature $T$, the Helmholtz
free-energy $F(V,T)$ per atom can be written as
\begin{equation}
F(V,T)=E_{c}(V)+F_{ion}(V,T)+F_{el}(V,T), \tag{1}%
\end{equation}
where $E_{c}$ represents the 0-K total energy which is obtained from
\textit{ab initio} electronic structure calculations, $F_{el\text{ }}$is the
free energy due to the thermal excitation of electrons, and $F_{ion}$ is the
ionic vibrational free energy$\ $which is evaluated from the partition
function $Z_{ion}$=$\exp(-NF_{ion}/k_{B}T)$. Here $N$ is the total number of
lattice ions. In the mean-field approximation, the classical $Z_{ion}$ is
given by \cite{Wasserman}
\begin{equation}
Z_{ion}=\left(  \frac{mk_{B}T}{2\pi\hbar^{2}}\right)  ^{3N/2}\left(  \int
\exp\left(  -g(\mathbf{r},V\right)  /k_{B}T)d\mathbf{r}\right)  ^{N}. \tag{2}%
\end{equation}
The essential of the MFP approach is that the mean-field potential
$g(\mathbf{r},V)$ is simply constructed in terms of $E_{c}$ as follows
\cite{Wang}
\begin{equation}
g(r,V)=\frac{1}{2}\left[  E_{c}(R+r)+E_{c}(R-r)-2E_{c}(R)\right]  , \tag{3}%
\end{equation}
where $r$ represents the distance that the lattice ion deviates from its
equilibrium position $R$. It should be mentioned that the well-known
Dugdale-MacDonald expression \cite{Dug} for the Gr\"{u}neisen parameter can be
derived by expanding $g(r,V)$ to order $r^{2}$. Then, $F_{ion}$ can be
formulated as
\begin{equation}
F_{ion}(V,T)=-k_{B}T\left(  \frac{3}{2}\ln\frac{mk_{B}T}{2\pi\hbar^{2}}+\ln
v_{f}(V,T)\right)  , \tag{4}%
\end{equation}
with
\begin{equation}
v_{f}(V,T)=4\pi%
{\displaystyle\int}
\exp\left(  -\frac{g(r,V)}{k_{B}T}\right)  r^{2}dr. \tag{5}%
\end{equation}

When the electron-phonon interaction and the magnetic contribution are
neglected, the electronic contribution to the free energy is $F_{el}$%
=$E_{el}-TS_{el}$, where the bare electronic entropy $S_{el}$ takes the form
\cite{Jarlborg}
\begin{equation}
S_{el}(V,T)=-k_{B}%
{\displaystyle\int}
n(\epsilon,V)\left[  f\ln f+(1-f)\ln(1-f)\right]  d\epsilon, \tag{6}\label{e1}%
\end{equation}
where $n(\epsilon,V)$ is the electronic density of states (DOS) and $f$ is the
Fermi distribution function. With respect to Eq. (6), the energy $E_{el}$ due
to electron excitations can be expressed as
\begin{equation}
E_{el}(V,T)=\int n(\epsilon,V)f\epsilon d\epsilon-\int^{\epsilon_{F}%
}n(\epsilon,V)\epsilon d\epsilon, \tag{7}%
\end{equation}
where $\epsilon_{F}$ is the Fermi energy. Given the Helmholtz free-energy
$F(V,T)$, the other thermodynamic functions such as the entropy $S=-(\partial
F/\partial T)_{V}$, the internal energy $E$=$F+TS$, the pressure
$P$=$-(\partial F/\partial V)_{T}$, and the Gibbs free energy $G$=$F+PV$, can
be readily calculated.

In all of our calculations we take the molybdenum structure to be
body-centered-cubic (bcc) structure (\textit{Im3m}). The 0-K total energy was
calculated using the full-potential linearized augmented plane wave (LAPW)
method \cite{Blaha} in the generalized gradient approximation (GGA)
\cite{Perdew}. We used constant muffin-tin radii of 2.05$a_{0}$ ($a_{0}$ is
the Bohr radius). The plane wave cutoff $K_{cut}$ is determined from
$R_{mt}\times K_{cut}$=10.0. $4000$ $k$ points in the full zone are used for
reciprocal-space integrations. The calculations were performed for atomic
lattice parameter ranging from 4.8$a_{0}$ to 6.8$a_{0}$. For the highest
calculated pressure the atomic volume has exceeded the touching sphere limit,
an extrapolation for getting the 0-K total energy points by Morse function has
been done.

The \textit{P-V} Hugoniot was obtained from the Rankine-Hugoniot
relation $P(V_{0}^{p}-V)/2$=$E-E_{0}^{p}$ for internal energy $E$,
pressure $P$, and volume $V$, which are achieved by shock from
initial conditions $V_{0}^{p}$ and $E_{0}^{p}$ for the present
porous Mo. Usually, the density of porous materials of different
porosities ($m<2\thicksim3$) becomes at once the same as that of
crystal under about 1.0 GPa shock-wave compression. In this case, it
is a good approximation to take the initial internal energy
$E_{0}^{p}$ for porous Mo to be exactly the same as the internal
energy $E_{0}$ for crystal Mo. The $V_{0}^{p}$ is
$V_{0}^{p}$=$mV_{0}$, where $m$ is the initial porosity and $V_{0}$
is the ambient volume of nonporous single-crystal Mo. Note that
during the calculations, we have neglected shock melting and the
phase dependence of the high-temperature equation of state, which is
in accord with the finding of Mitchell \textit{et al}.
\cite{Mit1991} that the effects of shock melting on the $P$-$V$
Hugoniots of several reference metals were too weak to be observed.

\begin{figure}[ptb]
\begin{center}
\includegraphics[width=1.0\linewidth]{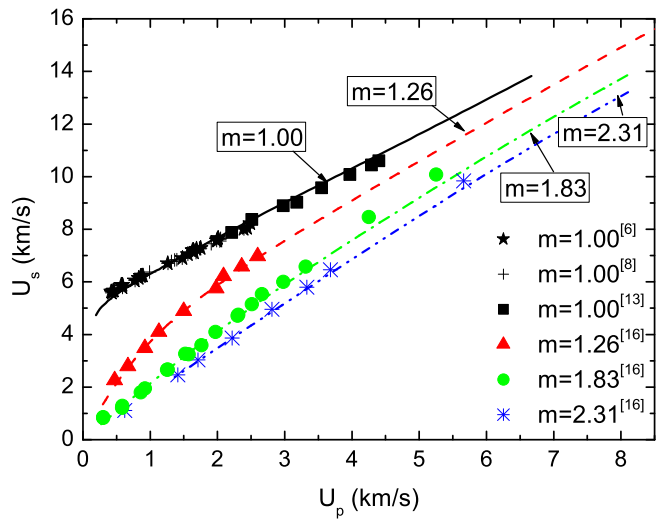}
\end{center}
\caption{(Color online) Calculated shock-wave velocity versus particle
velocity of molybdenum for initial porosities $m$=1.0, 1.26, 1.83, and 2.31.
Different calculated curves represent different porosities. For comparison,
the experimental data for $m$=1.0 (McQueen \textit{et al}.\textit{
}\cite{McQueen1}, LASL \cite{Marsh}, and Hixson \textit{et al}.\textit{
}\cite{Hixson}), 1.26, 1.83, and $2.31$ (Trunin \textit{et al}.\textit{
}\cite{Trunin1}) are also shown in the figure. }%
\end{figure}

Figure 1 shows the calculated results of $U_{p}$-$U_{s}$ diagram for $m$=1.0,
1.26, 1.83, and 2.31. Here $U_{p}$ is particle velocity and $U_{s}$ is
shock-wave velocity. For comparison, the experimental data
\cite{McQueen1,Marsh,Hixson,Trunin1} are also summarized in Fig. 1. For
$m$=1.0 (crystal Mo), our calculated $U_{p}$-$U_{s}$ Hugoniot agrees well with
the three groups of experimental data. Of these three groups of experimental
data, the data points of McQueen \textit{et al. }\cite{McQueen1} are obtained
by using explosive system. The data points Marsh \textit{et al. }\cite{Marsh}
(LASL Shock Hugoniot Data) are in the same particle velocity range from 0.4 to
2.5 km/s. The data points of Hixson \textit{et al. }\cite{Hixson} for $m$=1.0
are obtained by using the two-stage light-gas gun facility at Los Alamos
National Laboratory (LANL) with $U_{p}$ ranging from 2.2 to 4.4 km/s. The
least-squares fit to these three groups of data gives the relation $U_{s}%
$=5.109$+$1.247$U_{p}$ at $m$=1.0, while our calculated results gives $U_{s}%
$=4.915$+$1.347$U_{p}$. For $m$=1.26, it reveals in Fig. 1 that our calculated
$U_{s}$ versus $U_{p}$ relation agrees with the data points of Trunin
\textit{et al. }\cite{Trunin1} very well. For the two larger porosities
$m$=1.83 and $m$=2.31, our results agree well with the data points of Trunin
\textit{et al. }\cite{Trunin1} up to the particle velocity of $3.5$ km/s. At
even more higher particle velocity ($U_{p}>$4.0 km/s), the scarce attainable
experimental data for $m$=1.83 and $m$=2.31 show that the amplitude of the
shock-wave velocity is underestimated by our calculations, and this deviation
seems to be growing with increasing the particle velocity.

\begin{figure}[ptb]
\begin{center}
\includegraphics[width=1.0\linewidth]{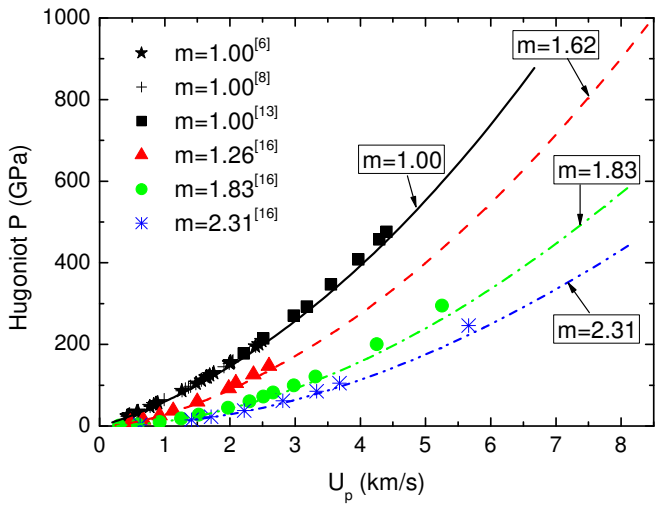}
\end{center}
\caption{(Color online) Calculated Hugoniot pressure versus particle velocity
of molybdenum for initial porosities $m$=1.0, 1.26, 1.83, and 2.31. The
experimental data points from several groups are also shown in the same way as
occurred in Fig. 1. }%
\end{figure}

Figure 2 shows the calculated Hugoniot pressures as functions of the particle
velocity $U_{p}$ for different porosities. The experimental data
\cite{McQueen1,Marsh,Hixson,Trunin1} are also plotted for comparison. For
crystal Mo at equilibrium, i.e., $m$=1.0, one can see that the theory agrees
well with the experiments over the full range of data. For $m$=1.26, our
calculated results are also in excellent agreement with the experiment by
Trunin \textit{et al}. \cite{Trunin1}. Note that for this porosity, there
still lacks the data points at high pressure, and our calculation above the
pressure of 147 GPa needs to be experimentally verified in the future. For
$m$=1.83 and 2.31, our results agree well with the experimental data points up
to the Hugoniot pressure of 100 GPa. With further increase of the particle
velocity, a slight deviation of the present calculated Hugoniot pressure from
the measurement occurs, with the latter systematically larger than the former
at given $U_{p}$.

Figure 3 shows our results of pressure dependence of Hugoniot volume (scaled
by ambient atomic volume $V_{0}$ of the crystal Mo) for different porosities.
Again, the attainable experimental data \cite{McQueen1,Marsh,Hixson,Trunin1}
are also illustrated in Fig. 3 for comparison. For $m$=1.0 and 1.26, the
agreement between our calculated results and experimental measurement is
obviously good. For the porosities $m$=1.83 and 2.31, on the other side, the
experimental points become scattered and are difficult to fit in a smooth
curve. For this reason, the difference between our calculated $V$-$P$ relation
and the experimental data is somewhat enlarged compared to the $U_{p}$-$U_{s}$
and $U_{p}$-$P$ relations. The scattering of data points is most likely due to
the fact that it is more difficult for larger porosity to prepare the samples
with the same initial porosity for several experiments. It should be mentioned
that the $V$-$P$ relation is more sensitive to the difference between
theoretical results and experimental points than that of $U_{p}$-$U_{s}$ and
$U_{p}$-$P$ relations.\begin{figure}[ptb]
\begin{center}
\includegraphics[width=1.0\linewidth]{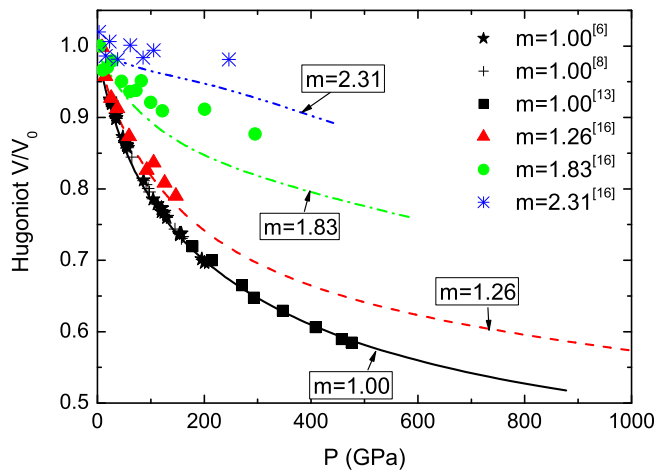}
\end{center}
\caption{(Color online) Calculated Hugoniot pressure versus relative atomic
volume of molybdenum for initial porosities $m$=1.0, 1.26, 1.83, and 2.31. The
experimental data points from several groups are also shown in the same way as
occurred in Fig. 1. }%
\label{fig3}%
\end{figure}

The calculated Hugoniot temperature is shown in Fig. 4 as a function of
pressure $P$ for several values of porosity. For comparison, the previous
calculations by McQueen \textit{et al. }\cite{McQueen} and Hixson \textit{et
al. }\cite{Hixson} for the case of $m$=1.0 are also shown in Fig. 4. Clearly,
our results for $m$=1.0 is in good agreement with the previous results. One
can see from Fig. 4 that the larger the initial porosity is, the steeper the
temperature curve is. This is due to that the porous material can absorb and
transform to heat the energy of shock-wave in the process of shock
compression.\begin{figure}[ptb]
\begin{center}
\includegraphics[width=1.0\linewidth]{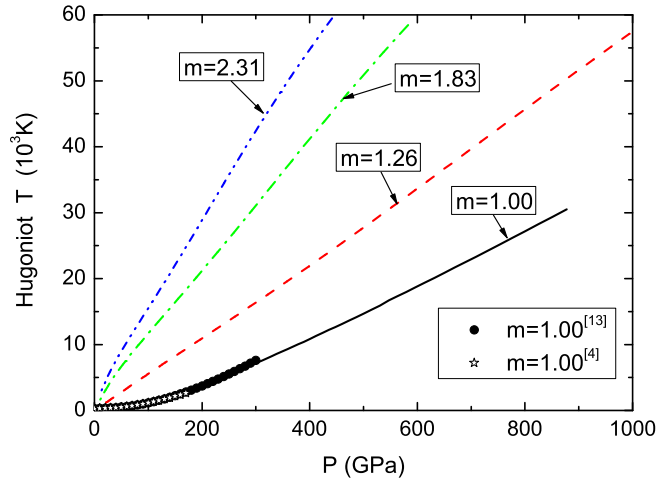}
\end{center}
\caption{(Color online) Calculated Hugoniot temperatures as functions of
Hugoniot pressure $P$ for molybdenum with initial porosities $m$=1.0, 1.26,
1.83, and 2.31. The filled circle and the star points represent theoretical
results of McQueen \textit{et al}.\textit{ }\cite{McQueen} and Hixson
\textit{et al}.\textit{ }\cite{Hixson}, respectively.}%
\label{fig4}%
\end{figure}

\begin{figure}[ptb]
\begin{center}
\includegraphics[width=1.0\linewidth]{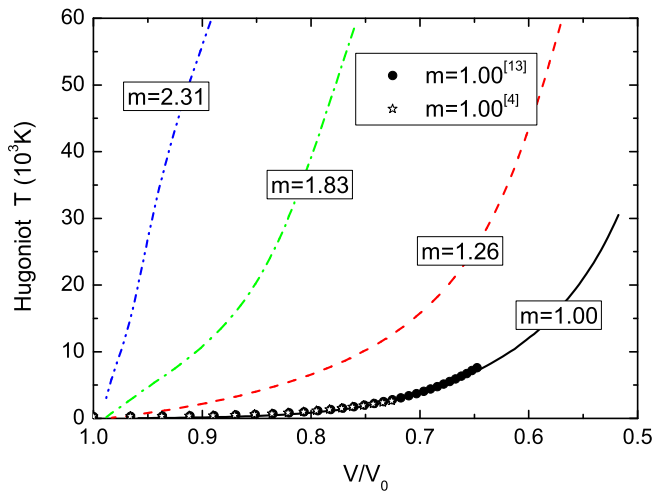}
\end{center}
\caption{(Color online) Hugoniot temperature versus relative volume of
molybdenum for initial porosities $m$=1.0, 1.26, 1.83, and 2.31. The filled
circle and star points represent theoretical results of McQueen \textit{et
al}.\textit{ }\cite{McQueen} and Hixson \textit{et al}.\textit{ }%
\cite{Hixson}, respectively.}%
\end{figure}

Finally, figure 5 shows the calculated Hugoniot temperature as a function of
atomic volume for the values of $m$=1.00, 1.26, 1.83, and 2.31. The previous
theoretical results \cite{McQueen,Hixson} for $m$=1.0 are also plotted in Fig.
5 for comparison. The agreement between our results and those of the two
groups \cite{McQueen,Hixson} is very good. Due to the reason mentioned above,
the Hugoniot temperature increases rapidly with the increase of initial
porosity at a given relative volume of Mo.

In summary, the Hugoniots of Mo with porosities $m$=1.0, 1.26, 1.83,
and 2.31 have been calculated by using the first-principles MFP
approach. Our results show good agreement with the experimental data
at $m$=1.00 and 1.26. For larger porosities of $m$=1.83 and 2.31,
the difference between our results and data points grows with the
pressure. The Hugoniot temperature of porous Mo have also been
calculated. For $m$=1.0, our calculated results are remarkably
consistent with the previous calculations. At present, there are no
measurement of Hugoniot temperature attainable for porous Mo to
confirm our theoretical results, and we leave this verification for
the future shock-wave experiments.

This work was partially supported by NSFC under grant number 10604010 and 10544004.

\end{document}